\documentclass[aps,prb,twocolumn,superscriptaddress]{revtex4}
\usepackage{graphicx}
\newcommand{\be}{\begin{eqnarray}}
\newcommand{\ee}{\end{eqnarray}}

\newcommand{\ds}[1]{#1{\hskip-2.2mm}/}

\begin{document}

\title{Effective Field Theory for the Quantum Electrodynamics of a Graphene Wire}
\author{P.~Faccioli}
\email{faccioli@science.unitn.it}
\author{ E.~Lipparini}
\email{lipparin@science.unitn.it}
\affiliation{Dipartimento di Fisica  Universit\'a degli Studi di Trento e I.N.F.N, Via Sommarive 14, Povo (Trento), I-38050 Italy.}
\begin{abstract}
We study the low-energy quantum electrodynamics of electrons and holes, in a thin graphene wire. 
We develop an effective field theory (EFT) based on an expansion in $p/p_T$, where $p_T$ is the typical momentum of  electrons and holes in the transverse direction, while $p$ are the  momenta
in the longitudinal direction. 
We show that, to the lowest-order in $(p/p_T)$, our EFT theory is formally equivalent to the exactly solvable Schwinger model.
By exploiting such an analogy, we find that the ground state of the quantum wire contains a condensate of electron-hole pairs. The excitation spectrum is saturated by  
electron-hole collective bound-states, and we calculate the dispersion law of such modes.
We also compute the DC conductivity per unit length at zero chemical potential  and find  $g_s \frac{e^2}{h}$, where $g_s=4$ is the degeneracy factor.
\end{abstract}
\maketitle

\flushbottom
\maketitle
\section{Introduction}
Monolayer graphene is the newest two-dimensional  electronic
system which has been fabricated in the last years. This system has
attracted an enormous
interest both from the experimental and theoretical sides, since
its electrons behave as Dirac fermions and show an unusual
sequence of Landau levels yielding to an
anomalous quantum Hall effect\cite{Nov05,Zha05}.

These features of graphene are based on a tight-binding 
model
calculation of its energy bands\cite{Ha88,Zhe02,Per06}, which shows a
nearly linear dispersion
of the energy bands close to the K point.
Information on the energy dispersionÄ of the Dirac cones
has been achieved by means of Raman spectroscopy\cite{Fer06,Gra07,Yan07,Pis07}.

Another important property of monolayer graphene is the possibility
of cutting out samples of desired form and size. Of particular interest among 
these
are quantum wires of graphene, which have been previously considered in several
approaches\cite{Nov07,Sil07,Per07}. In these studies, the properties of  the quantum wires 
have been investigated theoretically for different types of confinement and of boundary
conditions at the edge of the strip, neglecting the interaction among the
carriers. 

In this  work, we study the consequences of the electromagnetic interactions between electrons and holes, inside a wire made from 
a single layer of two-dimensional graphene, at zero chemical potential.  We show that the combined effects of the interaction
and of the low-dimensionality of the system drastically change the structure of the ground-state and of the spectrum of excitations,
which are typical of the two-dimensional graphene. In fact, the vacuum develops a condensate of electron-hole pairs and 
the spectrum of excitations turns out to be saturated by bosonic particle-hole collective modes.

Let  $v$ be the band velocity, $a$ be the lattice spacing  and, $L$ and $\lambda$ be  the length and the width of the wire, respectively. 
 In  graphene, $v \simeq 10^6 m ~s^{-1} \simeq 1/300~ c$, and the lattice spacing  $a\simeq~0.25~nm$.
We consider a system for which $a\ll \lambda, L$,  so  one can adopt a continuous formulation. In addition, we assume $\lambda\ll L$ so that 
the energy associated to longitudinal momenta   $E_L=\hbar v/L$ 
decouples from the energy associated to transverse momenta, $E_T=\hbar v/\lambda$.

Our goal is to exploit such a separation of scales to build an EFT, in which the explicit low-energy degrees of freedom 
are the electrons and holes which propagate in the longitudinal direction, and interact by exchanging quanta of the electromagnetic field.
We shall show that the interactions drastically change the properties of the wire: the spectrum gets saturated by  
electron-hole collective bound-states. The electric conductivity turns out to be ${e^2\over h}$ in units of the degeneracy factor $g_s=4$, 
i.e. the same result of Landauer's formula for the conductivity of a classical one-dimensional wire\cite{Lan70}.

The paper is organized as follows. In section \ref{free} we introduce the model  Hamiltonian  describing the motion of electron and holes without taking into account of electrostatic interactions between charge carriers.
In section \ref{reviewEFT} we briefly review the main ideas of EFT and we construct the lowest order Lagrangian, in which the effect of Coulomb interactions is systematically taken into account. In section \ref{sch} we show 
that, under suitable 
approximations, such a Lagrangian reduces to the Schwinger model.
In section \ref{solu} we shall review the exact solution of such a theory. The implications of these results on graphene physics will be presented in section \ref{implications}. Conclusions and perspective developments
are summarized in section \ref{conclusions}.

\section{Theory for graphene wires without Coulomb interactions}
\label{free}
  
We begin by considering the free quantum motion near the K point of the electrons in the conductance band and holes in the valence band of graphene. The corresponding second-quantized Hamiltonian leading to the desired single-particle spectrum
\be
\omega(k)=  \pm \hbar v | k|,
\label{disp}
\ee
is of course
\be
\label{H0}
H_0 = v \hbar \int d x ~ \psi^\dagger (t,x)\left( -i \sigma_1 \partial_x \right) \psi(t,x),
\label{Ham}
\ee
where $\psi(t,x)$ is the fermion field operator. In the following, we shall work in the "natural units" of this problem, in which 
 $\hbar=v=1$. Note that, in such units, the speed of light is  $c=1/\beta\simeq 300$.

In order to exploit the formal analogy with the relativistic Dirac theory, it is convenient to introduce  position contravariant  vectors 
\be
\tilde x^\mu=(v t, x) = ( t, x),
\ee  
momentum contravariant vectors  
\be
\tilde p^{\mu}=(E/v, p)=(E,p),
\ee  
and the metric tensor as $g_{\mu \nu}=\textrm{diag}[1,-1]$. 
In addition, let $\gamma^\mu$ and $\gamma^S$ be  $2\times 2$ matrices obeying the usual Dirac algebra:  
\be
\{\gamma^\mu, \gamma^\nu\} &=& 2  g^{\mu \nu}\\
  \{\gamma^S, \gamma^\mu\} &=& 0
 \ee
and, $(\gamma^S)^2 =  1$.
For example, one may choose a representation in which
\be
\gamma^0 &=&  \sigma_3\nonumber\\
\gamma^1 &=& i \sigma_2 \nonumber\\
\gamma^S &=&  \gamma^0 \gamma^1= \sigma_1 
\label{gamma}
\ee
Note that with this choice  one has $\gamma^\mu \gamma^S =  - \epsilon^{\mu \nu} \gamma_\nu.$

Using such a set of definitions, the action associated to the free "Dirac" Hamiltonian (\ref{Ham}) can be cast in the familiar form
\be
S_0 =   \int d^2 \tilde x ~  \bar{\psi} ~i ~ \tilde {\ds\partial} ~ \psi.
\ee

The purpose of this work is to study how the properties of the free "Dirac" particles described by 
Hamiltonian (\ref{Ham}) are modified once electromagnetic 
interactions are systematically introduced. The construction of the Lagrangian for the theory with electromagnetic interaction
will be presented in the next section, using the framework of EFT.

\section{Effective field theories}
\label{reviewEFT}

Effective Field Theories are systematic  low-energy approximations of  more microscopic theories  ---for a pedagogic introduction see  e.g.  \cite{howto}---. 
The main idea underlying the EFT formalism  is  the familiar observation that  the low-energy processes are insensitive to the \emph{details} of the ultra-violet physics.  
For example, the long wavelength classical radiation generated by  complicated current source ${\bf J}({\bf r},t)$  
of size $d$ can be replaced by a sum of point-like multi-pole currents $E1, M1, \ldots$. The usefulness of such an expansion is that, for  wavelengths $\lambda \gg d$,  
usually only few multipoles are usually needed for sufficient accuracy.
%
The  gain in going from the fundamental theory to its  low-energy  EFT resides in the fact that  the latter is in general much simpler,  as it describes the dynamics of  fewer degrees of freedom.  
Whenever an EFT can be built, all predictions for low-energy observables of  the underlying microscopic theory can be reproduced to a desired level of accuracy, truncating the expansion at a finite order. 
However, the price to pay for such a simplification is that of having to specify a finite number of  unknown parameters. These have to be determined
 from experiment or from microscopic calculations in the underlying  (more) fundamental theory.

The transformation from a microscopic theory to a simpler EFT, valid  for low-momentum processes, is achieved in two steps. 
The first step consists in introducing  a hard cut-off  of order the momentum at which the physics we are not interested in describing 
 becomes important. Only momenta lower than such a cut-off are retained in calculations.  
The second step consists in writing the Lagrangian or the Hamiltonian for the low-energy degrees of freedom.  
According to Weinberg theorem\cite{Weinberg}, in quantum field theory this task can be achieved starting from  the most general Lagrangian, compatible
 with the symmetry properties of the underlying fundamental theory and with the fundamental principles of quantum field theory,
\be
\label{EFT}
\mathcal{L} = \sum_{i=1}^{\infty} c_i(\Lambda) ~\hat{O}_i.
\ee 
Note that  the effective Lagrangian $\mathcal{L}$ contains infinitely many operators $O_i$, along with an infinite number of corresponding unknown effective coefficients $c_i$. 
Hence, the expression (\ref{EFT})  has no predictive power, thus does not yet represent a physical theory. The usefulness of the EFT scheme resides in the possibility of establishing a power-counting scheme, i.e. a 
prescription to retain only a finite number of terms in the effective lagrangian (\ref{EFT}).
This possibility is based on the observation that the contribution of the operators with higher and higher dimensionality  in (\ref{EFT}) to any 
Green's function with external momenta of order $p$ is suppressed by powers of the ratio of momentum over the cut-off, i.e. $p/\Lambda$.
Hence,   predictions to an arbitrary level of accuracy can be obtained by retaining only a {\it finite} number of lowest-dimensional operators in (\ref{EFT}).
The corresponding effective coefficients have to be determined from microscopic calculations in the underlying fundamental theory, or by performing a {\it finite} number of experiments.

As an example, let us consider the EFT for the ordinary QED in vacuum, for momenta much below a 
cut-off scale $\Lambda$, chosen above the electron mass. 
In 3+1 dimension and natural units  ($\hbar=c=1$) the electron field and the photon field have mass dimension $3/2$ and $1$, respectively, while the electric charge is dimensionless. 
Hence, the lowest-dimensional operators which appear in the most general Lagrangian compatible with Lorentz and gauge symmetry are:
\be
\label{QEDlowk}
\mathcal{L}^{(\Lambda)} &=& \bar \psi\,( i \ds \partial - e(\Lambda) A \cdot \gamma - m(\Lambda) )~ \,\psi
-\frac{1}{4} F_{\mu \nu} F^{\mu \nu}\nonumber\\
&+& e(\Lambda) c_1(\Lambda)~ \bar \psi\, \sigma_{\mu \nu} F^{\mu \nu} \psi \nonumber\\
&+& e(\Lambda) c_2(\Lambda)~ \bar \psi\, i \partial_\mu F^{\mu \nu} \gamma_\nu \psi 
+ d_{2} (\bar \psi \gamma_\mu \psi)^2 \nonumber\\
&+& \ldots
\ee
Some comments on this expression are in order. First of all, we note that the second and third lines  display  terms  which are not present in the original QED Lagrangian.  
The role of such effective local  interactions is to mimic the ultra-violet physics  above the cut-off $\Lambda$.   Note also that the operators in different lines have different mass dimensions: the operators in the 
first line have dimension $4$, that in the second
line has dimension $5$ and those in the last line have dimension $6$.  Since in natural units the action must be dimensionless, each term in  the Lagrangian must have mass dimension 
equal to the number of space-time dimensions. 
This condition  fixes the mass dimension of the effective coefficients to $[c_1]=-1$, $[c_2]=-2$ and
$[d_2] = -2$. 

A simple dimensional analysis implies that, if the external momenta $p's$ are small compared to the cut-off, 
 then the contribution to the Green's functions of the terms in the second  and third lines of Eq.~(\ref{QEDlowk}) are suppressed by increasing powers of $(p/\Lambda)$, with respect to the contribution of the
 terms in the first line. Hence, if the cut-off is sent to inifinity, only the first line provides finite contributions. For this reason, in the language of effective field theory, the operators with mass dimension larger than the number of space-time dimensions are called \emph{irrelevant}. 
 
 Finally, we observe also that effective field theories are in general non-renormalizable. However, 
 that does not lead to problems, because  the cut-off $\Lambda$ is always kept finite. 
The effective field theory (\ref{QEDlowk}) retains predictive power, at the price of computing the coefficients $c_1(\Lambda), c_2(\Lambda)$  and $d_2(\Lambda)$ microscopically from 
QED, or to fix them from experimental measurements.

Let us now carry out the same program in our specific case, in which the cut-off scale is provided by the transverse momentum in the wire, $\Lambda=p_T$. 
We are interested in an effective field theory which describes only the dynamics of the low-energy degrees of freedom inside the wire, i.e. 
the electrons, holes and photons propagating along the longitudinal direction. On the other hand, we are not interested in describing the dynamics of the on-shell photons, which are radiated away from the wire.
Hence, it is sufficient to consider an effective field theory in 1+1 dimensions,  in which a  \emph{dynamical} two-component  photon effective field is introduced to mediate the electron-hole interaction, inside the wire.
In addition, the  electrons and holes are allowed to interact with an \emph{external} electro-magnetic field, which is therefore defined in $3+1$ dimensions. 
 
The starting point to construct an effective Lagrangian which satisfies these requirements is to introduce a gauge invariant coupling of photons and fermions:
\be
\tilde D_\mu=\tilde{\partial}_\mu + i e~\tilde{a}_\mu + i e \tilde A^\textrm{ext}_{\mu},
\ee
where we have used the following definitions:
\be
\tilde{a}_0 &=& a_0,\qquad \tilde{A}^\textrm{ext}_0 = A^\textrm{ext}_0\\
\tilde{a}_1 &=& \beta ~a_1,\quad \tilde{A}^\textrm{ext}_1= \beta~ A^\textrm{ext}_1.
\ee
$\tilde a_\mu$ is the quantized dynamical field which describes the photons inside the wire, while $A^\textrm{ext}_\mu$ is a classical external field. 
Notice that, in $1+1$ dimensions, the fermion field has mass dimension $1/2$, while the $a_\mu$ and $A_\mu$ field have mass dimension $1$.
The  lowest-dimensional gauge invariant and "Lorentz"-invariant operators are
\be
&&\bar \psi\, i \tilde{\ds D} \psi, ~m(\Lambda) \bar \psi \psi,~ (\bar \psi O \psi)^2,   \qquad( O=1, i \gamma_s, \tilde \gamma_\mu,\tilde \gamma_\mu \gamma_S) \nonumber\\
&&  \bar \psi  \sigma_{\mu \nu} F^{\mu \nu}  \psi \nonumber\\
&& f_{\mu \nu} f^{\mu \nu},~  \bar \psi\, i \tilde \partial_\mu F^{\mu \nu} \gamma_\nu \psi.
\ee
The first, second and third lines contain operators of dimension 2, 3 and 4, respectively.
Some of such terms can be rejected,  based on physical considerations. First of all,  we want our  theory to reproduce the free theory results, in the limit in which  the electromagnetic 
coupling is set to 0. In order to ensure such a condition,  we must  not
include the mass term, which would spoil the linearity of the fermion dispersion relation, near the K-point.  

The term $\bar \psi  \sigma_{\mu \nu} F^{\mu \nu}  \psi$ describes in general the interaction of the electro-magnetic field with the magnetic moment of the fermion field. 
However, one should  keep in mind that the Pauli matrices in the "Dirac-like" theory for graphene do not have the real physical interpretation of spin matrices. Clearly,  an interaction between 
magnetic  field and pseudo-spin would not be physical. 

The term   $\bar \psi\, i \tilde \partial_\mu F^{\mu \nu} \gamma_\nu \psi$ can be eliminated using the Eq. of motion for the $a_\mu$ field
generated by the terms which are  present at this order. Hence, its only effect is that of rescaling the existing effective coefficients.

The contact terms   $ (\bar \psi O \psi)^2$ deserve a particular attention. In principle, these terms appear already at the lowest-order in our  effective  Lagrangian and should be kept into
account. Their role is to mimic the ultra-violet physics which sets in when electrons and holes interact at a distance of the order of the inverse cut-off, i.e. of the transverse size of the wire, 
$\simeq  1/p_T$. 
The inclusion of such terms could in principle be dealt in the context of the massless Thirring model~\cite{zinn}.  
However, in the limit of extremely small electron density we are considering, the correlations brought in by the contact interactions are certainly small, compared to the
the long-ranged Coulombic interaction, provided by the $\bar \psi\, i \tilde{\ds D} \psi$ and $  f_{\mu \nu} f^{\mu \nu}$ terms. Hence, we chose to  neglect the effects of the contact terms in our first work. 
Finally, we observe that a finite electron density  would give raise to a term $\mu \bar \psi \gamma_0 \psi$, which would appear at the leading-order, as expected.  

Strictly speaking, some of the approximations made  above could spoil the power-counting scheme of our effective field theory. This means that our results may retain some leading-order dependence on the specific 
choice of the cut-off. However, in our specific case, this does  not represent a problem. In fact, unlike the example of low-energy QED, the choice of the cut-off is not arbitrary, but it is 
determined  by the geometry of the wire, i.e. by its transverse size.  

In conclusion, our lowest-order effective action for the internal quantum electrodynamics of the wire reads
\be
\label{S}
S \equiv  \int d^2\tilde x ~\left(\bar \psi\, i \tilde{\ds D} \,\psi -  \frac{c(\Lambda)}{4} f_{\mu \nu} f^{\mu \nu}~\right),
\ee
where $f_{\mu \nu} = \tilde{\partial}_\mu \tilde a_\nu - \tilde{\partial}_\nu \tilde a_\mu =  \partial_\mu a_\nu - \partial_\nu a_\mu$.

\section{Mapping the Electrodynamics of the wire onto the Schwinger Model} 
\label{sch}
   
The present work is driven by the observation that effective action (\ref{S}) can be formally mapped onto QED with massless fermions in 1+1 
dimensions, i.e. in the celebrated  chiral {\it Schwinger model}~\cite{schwref}.  This is an  exactly solvable model which exhibits an extremely rich non-perturbative dynamics --- for a detailed discussion see, e.g., 
\cite{schwref, casher} and references therein---.
To establish such a connection it is sufficient to re-absorbe the effective coefficient $c(\Lambda)$  in the definition of the photon field, by setting  
$\sqrt{c(\Lambda)} \tilde a_\mu~\to~\tilde a_\mu$. 
Once this has been done,  a factor $1/\sqrt{c(\Lambda)}$  appears in the electromagnetic coupling operator between the fermions and the rescaled photon field. 
In conclusion,   at the leading-order in the expansion parameter  $(p \lambda)$, the quantum electrodynamics of electrons and holes inside the wire is described by the effective action
\be
\label{Ssh}
S= \int d^2 \tilde x \left[~\bar \psi\, i \tilde{\ds D} \,\psi -  \frac{1}{4} f_{\mu \nu} f^{\mu \nu}\right],\quad \mu,\nu=0,1
\ee
where the covariant derivative is defined as 
\be
\label{cov}
\tilde D_\mu=\tilde{\partial}_\mu + i g~\tilde{a}_\mu + i e \tilde A^\textrm{ext}_{\mu},
\ee
 with $g=\frac{e}{\sqrt{c(\Lambda)}}$. 
 
 Eq.(\ref{Ssh}) defines the action of the Schwinger model.
We stress the fact that  factor $1/\sqrt{c(\Lambda)}$ appears only in the coupling of the fermions with the {\it quantized} and rescaled electromagnetic field $\tilde a_\mu$ and 
not in the coupling with the classical external field $A^\textrm{ext}_\mu$. 
We also emphasize the fact that such a coefficient is not predicted in our approach. It has to be determined either from experiments, or from microscopic calculations.

\section{Exact Solution of the Schwinger Model}
\label{solu}
A remarkable feature of massless QED in 1+1 dimensions is that the variation of the action generated by the two independent rotations of the fermion fields
\be
\label{rotgauge}
\psi(x)&\to& e^{i \chi(x)} \psi(x)  \qquad  \bar \psi(x)\to \bar \psi(x) e^{-i \chi(x)}\\
\label{rotchiral}
\psi(x)&\to& e^{i \gamma^S \Phi(x)} \psi(x)  \quad  \bar \psi(x)\to \bar \psi(x) e^{i \gamma^S \Phi(x)}
\ee
can be re-absorbed by two local gauge transformations:
\be
\label{gauge}
\tilde{a}_\mu &\to& \tilde{a}_\mu - \frac{1}{g} \tilde{\partial}_\mu \chi,\\
\label{chiral}
\tilde{a}_\mu &\to& \tilde{a}_\mu + \frac{1}{g}~\epsilon_{\mu \nu} \tilde{\partial}^\nu \Phi.
\ee

The dynamical consequences of this fact become evident once one parametrizes the photon field degrees of freedom as
\be
\tilde{a}_\mu = \frac{1}{g} \left(\tilde{\partial}_\mu \chi - \epsilon_{\mu \nu} \tilde{\partial}^\nu \phi~\right)
\label{subs}
\ee
and  re-expresses the path integral in terms of the fermion fields $\psi, \bar \psi$ and of the $\chi$ and $\phi$ fields.
The $\chi$ field  is pure-gauge, hence it has a vanishing field strength tensor. In addition, the fermionic measure $\mathcal{D}\psi \mathcal{D} \bar \psi$  is left invariant by the gauge rotation (\ref{rotgauge}).
 Hence, the $\chi$ field is unphysical, and  can be completely eliminated from  the  path integral.
On the other hand, the $\phi$ field contributes to the  field strength tensor, through the term $2\frac{ 1}{g^2} ~  \tilde{\partial}^\mu \phi~  \tilde \partial^2~\tilde{\partial}_\mu~\phi$.
In addition, an anomalous term appears in the action, as a consequence of the fact that the chiral rotation (\ref{rotchiral})  does not leave the functional measure of the fermion fields  invariant: 
\be
\mathcal{D} \psi \mathcal{D} \bar \psi \to \mathcal{D} \psi \mathcal{D} \bar \psi \mathcal{J}^{-2}, 
\ee
where 
\be
\mathcal{J}^{-2}= \exp\left[-2 \int d^2 \tilde x~\frac{ 1}{g^2} ~  \tilde{\partial}^\mu \phi~  m_\phi^2 \tilde{\partial}_\mu~\phi\right]
\ee is the functional Jacobian determinant of the chiral transformation (\ref{rotchiral}) and $m_\phi=g/\sqrt{\pi}$ is the so-called Schwinger "mass". 

As a result, the path integral of the leading-order effective action reads
\be
Z= \int \mathcal{D}\psi  \mathcal{D}\bar \psi \mathcal{D} \phi ~\exp\left[ i S_\textrm{eff}[\psi, \bar \psi, \phi]\right],
\ee
where the effective action $S_\textrm{eff}$ is defined as
\be
\label{Seff}
&& S_\textrm{eff}[\psi, \bar \psi, \phi]  =
 \int d^2 \tilde x  ~\bar \psi\, i \tilde{\ds \partial} \,\psi   \nonumber\\
 &-&\frac 1 2 ~\frac{ 1}{g^2} ~  \tilde{\partial}^\mu \phi~ \Big( \tilde \partial^2- m_\phi^2 \Big)~\tilde{\partial}_\mu~\phi.
\ee

The effective action (\ref{Seff}) is quadratic in the both the fermion and boson fields. Thus, the path integral defining the corresponding quantum theory is Gaussian, so arbitrary $n-$point Green's functions
can be evaluated exactly.  All technical details of such calculations can be found in the original papers  \cite{schwref, casher} and in standard 
quantum field theory books ---such as e.g. \cite{zinn, Peskin}--- and  will not be repeated here. 

Let us begin by discuss the vacuum structure. The spectrum of scalar states is related to the poles of  the two-point Green's function:
\be
\label{G}
G(q)= \int d^2 \tilde x e^{i \tilde q\cdot \tilde x} \langle \Omega| T[\bar \psi(\tilde x) \psi(\tilde x) \psi(0)\psi(0)]|\Omega\rangle,
\ee
After performing the analytic continuation to the Euclidean space one finds (see e.g. \cite{zinn})
\be
\label{p2p}
 && \langle 0| T[\bar \psi(\tilde x) \psi(\tilde x) \psi(\tilde 0)\psi(0)]|0\rangle 
 \nonumber\\
 &=& \frac{K}{2 \pi^2 \tilde x^2} e^{-4 \pi (-\Delta(m_\phi, \tilde x) + \Delta(0,\tilde x))},
\ee
where $\Delta(m_\phi,\tilde x)$ is the Fourier transform of the $\phi$-field propagator:
\be
\Delta(m_\phi, \tilde x) = \int \frac{d^2\tilde p}{4 \pi^2}~\frac{1}{\tilde p^2 + m_\phi^2} ~e^{i \tilde p \cdot \tilde x}
\ee
and
\be
K= e^{-4 \pi (\Delta(m_\phi, 0) - \Delta(0,0))}.
\ee

It is immediate to verify that the point-to-point Green's function (\ref{p2p}) does not vanish at large Euclidean distances. This implies that the vacuum has a  non-trivial structure: it contains a finite density of
fermion-antifermion pairs
\be
\label{conde}
\langle \Omega|  \bar \psi \psi |\Omega \rangle= -\frac{g}{2 \pi \sqrt{\pi}} \exp({\gamma}),
 \ee
where $\gamma$ is the Euler constant.  Note that, in this expression, we have dropped space-time labels, since the ground state is static and invariant under translations along the longitudinal 
directions.  

Another remarkable feature of the Schwinger model is that all the zero-mass singularities of the Green's function (\ref{G})  exactly cancel out. The only singularity are free bosonic states, located at 
\be
\tilde q^2 = n^2 m_\phi^2, \qquad n=1,2,3, ..
\ee
The physical interpretation of this fact is that fermionic excitations are lifted from the spectrum, and only bosonic collective states exist.

This result can be proven as follows. First, we bosonize the free fermion fields $\bar \psi$ and $\psi$, introducing a free massless boson field $\theta$. The action becomes:
\be
\label{S2}
&& S[\theta, \phi]  =
 \int d^2 \tilde x \frac 1 2 (\tilde \partial_\mu \theta)^2   \nonumber\\
 &-&\frac 1 2 ~\frac{ 1}{g^2} ~  \tilde{\partial}^\mu \phi(x) \Big( \tilde \partial^2- m_\phi^2 \Big)~\tilde{\partial}_\mu~\phi(x).
\ee
Upon performing a translation of the $\theta$ field,
\be
\theta+ \phi/\sqrt{\pi}\to \theta
\ee
and integrating over $\phi$  one obtains a quantum field theory defined by the action
\be
\label{S0}
&& S[\theta]  =
 \int d^2 \tilde x \frac 1 2 (\tilde \partial_\mu \theta)^2   - \frac 12 m_\phi^2 \theta^2,
\ee
which corresponds to a free theory of scalar bosons.

\section{Physical Properties of the Graphene Wire with Coulomb interactions}
\label{implications}

Based on the mapping between the Schwinger Model and our EFT for the electrodynamics of the graphene wire,
 we can use the results listed in the previous section to obtain non-trivial predictions 
for the physical properties the system we are considering. 
In the following, we list some  most important predictions for the physical properties of the 
graphene wire, at zero temperature and chemical potential.

\subsection{ Ground-state structure}

The existence of a finite vacuum condensate, Eq. (\ref{conde}),  means that the ground-state of the quantum wire does not correspond to a configuration in which 
there is no electron in the conductance band and no hole in the valence band. Instead, 
 it  contains a finite density of scalar electrons-hole pairs.
 \begin{figure}[t!]
\includegraphics[width=8cm]{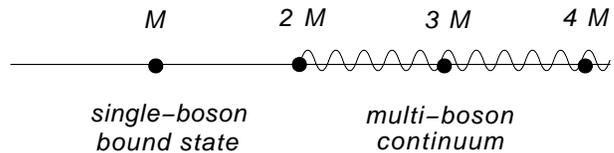}
\caption{Structure of the spectrum of collective excitations in the graphene wire.}
\label{fig1}
\end{figure}

\subsection{ Bosonization of the spectrum of excitations}

In the previous section,  we have seen that the spectrum of the Schwinger model does not contain single fermion excitations, 
but only arbitrary number of free fermion-antifermion bound-states (Schwinger bosons), of mass $M= g^2/\pi$ and spin-0.  
In the context of graphene theory, this feature implies that the spectrum of
 excitations of the wire starts with a single collective boson excitation, with dispersion relation 
\be
\label{spectrum}
\omega(k) = \pm \hbar v  \sqrt{ k^2 + \frac{4 \alpha/\beta}{c(\Lambda)}}
\ee
and contains a continuous of multi-boson excitations, starting at the two-boson threshold. Additional thresholds for multi-boson excitations are located at $n M$, with $n=3,4,5..$ ---see Fig. \ref{fig1}---.
In these formulas, we have restored the  $v$, $c$ and $\hbar$ constants and $\alpha={e^2\over4\pi\hbar c}\simeq1/137$ is the  fine structure  constant. 

The bosonization of the spectrum is a consequence of the fact that the theory is defined in 1+1 dimension and the fermions are massless. 
Indeed, it is observed also in  Luttinger liquids. 
However, an interesting feature of  the graphene wire is that  
 the bosonization occurs at vanishing chemical potential. Note that 
from the measurement of  the gap $\Delta = 2\sqrt{ \frac{\alpha/\beta}{c( \Lambda)} }$ it is in principle possible to 
 determine the leading-order effective coefficient $c(\Lambda)$ of our EFT.

\subsection{Conductivity of the graphene wire}

The electric conductivity of the graphene wire is defined as
\be
\Re\left[\sigma(\omega, q)\right] = \frac{1}{E^{ext}(\omega, q)} ~\Re \int d^2 
\tilde x e^{i (\tilde \omega \tilde{x}_0- \tilde q \tilde x)} \langle j_1(\tilde x) \rangle_{\tilde A^{\mu}},\nonumber\\
\ee
where $\tilde A^\mu(\tilde x)=(\Phi(\tilde x),0)$  
the potential of a weak external electric field, $E^\textrm{ext}$.

Applying the linear response theory, one immediately finds
\begin{equation}
\int d^2 \tilde x~ e^{i \tilde q\cdot \tilde x} ~\langle j^1(\tilde x)\rangle_{\tilde A^\mu_0} 
=  i~(i \Pi^{1 0}(\tilde q))~ \tilde A_0^{ext}(\tilde q),
\end{equation}
where $\Pi^{\mu \nu}(\tilde q)$ is the vacuum polarization tensor, defined as:
\be
i \Pi^{\mu \nu}(\tilde q) = \int d^2 \tilde x ~e^{i \tilde q \cdot \tilde x} 
~\langle \Omega| T[ j^\mu(\tilde x) j^\nu(0)] |\Omega \rangle.  
\ee
In the Schwinger model, this matrix elements can be computed exactly and reads
\be
\label{pol}
i \Pi^{\mu \nu}( \tilde q) =  - \left(g^{\mu \nu}- \frac{\tilde q^\mu \tilde q^\nu}{\tilde q^2}\right)~\frac{g^2} {\pi}
\ee
This result can be used to readily obtain the conductivity of the wire. 
After restoring the appropriate powers of $v$ and $\hbar$, we find our final results:
\be
\Re\left[{\sigma(\omega,  q)\over L}\right] &=& \Im 
{e^2\over\pi\hbar}{\omega/v\over(\omega/v+ q)(\omega/v- q)}
\label{cond}
\ee

and by taking the Fourier transform\cite{Kaw95}  
\be
\Re\left[{\sigma(\omega, x)\over L}\right]=\frac{e^2}{2\pi \hbar}\cos \left({\omega x\over v}\right)~.
\ee
Thus in the limit $\omega\to0$, we get for the DC conductance $G$ (defined as the $\omega\to0$ limit of 
$\Re\left[{\sigma(\omega, x)\over L}\right]$):
\be
G={e^2\over h}~.
\label{cond2}
\ee
This result must then be multiplied for the factor $g_s=4$, to account for the spin and sub-lattice degeneracy.
 \begin{figure}[t!]
\includegraphics[width=6cm]{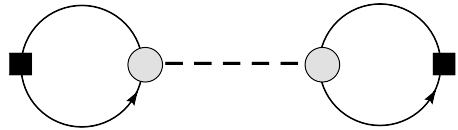}
\caption{Graphical representation of the  current-current correlation function $\Pi_{\mu \nu}$, in a graphene wire. 
The black square represents the electromagnetic current operator $J^\mu = \bar \psi \gamma^\mu \psi$, the  dashed line denotes the Schwinger boson propagator, 
while the grey circles represent the
 electron-hole wave-function in the Schwinger boson state. The induced current is  a consequence of quantum fluctuations of the Schwinger boson into electron-holes pairs. This Fig. was drawn using \it{Jaxodraw}~\cite{jaxo}.}
\label{fig2}
\end{figure}

A few comments on this result are in order. First of all, we note that the conductance
 does not depend on the effective coefficient, and therefore on the scale factor $\lambda$. 
Furthermore, we note that the current-current correlation function (\ref{pol})  is completely saturated by the pole corresponding to the Schwinger boson. 
This means that the correlation between the currents is mediated 
by the exchange of a massive composite bound-state.
The current induced by the external field is a purely quantum effect, mediated by the chiral anomaly. Physically, it arises from the polarization of the electron and holes constituents, inside the massive  Schwinger bosons ---see Fig. \ref{fig2}---.

Finally, we observe that 
the result (\ref{cond2}) for the conductance is in good agreement with a recent experimental observation of sub-band formation in graphene nano-ribbon~ \cite{exp}.

\section{Conclusions}
\label{conclusions}
In this work, we have presented the first study of the effects of the electric interactions between  electrons and holes, inside a thin graphene wire. To this end, we have 
developed EFT based on an expansion in $p/p_T$, and we have shown that the lowest-order of such a theory can be formally mapped onto the exactly solvable Schwinger model.
Using such an equivalence, we have shown that the electromagnetic interaction between electron and holes in a quasi one-dimensional 
wire dramatically affects  its longitudinal dispersion law, already at zero chemical potential. The spectrum of the wire 
contains only collective particle-hole excitations with dispersion characterized by a finite gap.
The DC conductance at zero chemical potential is given by  ${e^2\over h}$, in units of the degeneracy factor $g_s=4$.

It should be stressed that the present model is expected to work only in an appropriate range of length and witdh. In particular, if the width of the wire is too narrow, the shape
of the edge will significantly affect the electronic structure
and change the Dirac equation in the continuum approximation. On the other hand, if
the wire is too wide, the lowest-order  approximation of the present EFT would become insufficient. Hence, it would be interesting to compare the predictions of our model
with the results of lattice 
simulations for the electrodynamics two-dimensional graphene systems of different lengths and widths, using e.g. the techniques developed in~\cite{lattice}. On the one hand, this would provide a microscopic calculation of the unknown 
effective parameter $c(\Lambda)$. On the other hand, it would allow  to identify the region of transverse and longitudinal momenta, where the present EFT is applicable.
Another possible development of the present work would be to investigate how the properties of the wire change as a function of the fermion chemical potential and of the temperature.

\end{document}